\documentclass[prl,aps,twocolumn,superscriptaddress,amsmath,amssymb]{revtex4}
\begin{document}
\title{Reply to ``Comment on `Tunnel Window's Imprint on
Dipolar Field Distributions' ''}

\maketitle

Stamp and Tupitsyn (ST) have recently posted a Comment
\cite{ST} on a paper of ours \cite{ourPRL}. They claim \cite{ST}
that, in Ref. \cite{ourPRL}, ``Short-time relaxation was caused by a 
simple noise
field, acting uniformly over an `energy window' of width $2\delta 
h_{hf}$ (to simulate the
fluctuating nuclear spin bias)''. We see no basis for this statement.

Whereas previous work of Prokof'ev and Stamp \cite{PSPRL,PS1} deals 
with nuclear spin bath
mechanisms that give rise to a tunnel energy window (TEW) for 
single-molecule magnets,
such as Fe$_8$, the work we report in Ref. \cite{ourPRL} is independent of such
mechanisms. The only definition of the TEW, $\delta h_{hf}$, we make 
use of in Ref.
\cite{ourPRL} is the one we give therein in the abstract: ``we allow 
spin flips only if
the corresponding energy change is less than some $2\delta h_{hf}$''. 
This is further
explained below Eq. (2) in Ref. \cite{ourPRL}, where the system under 
consideration is
defined as an Ising model of spins (that clearly stand for the 
electronic cluster spins)
interacting through dipolar fields with the flipping rule just 
mentioned. Thus, the
definition of $\delta h_{hf}$ stands on its own without need for 
reference to the nuclear
bath.

In addition, we mention hyperfine interactions only
in passing, in an introductory remark, but the phrase ``typical 
hyperfine coupling'',
which PT ascribe to us, does not appear in Ref. \cite{ourPRL}. We do 
refer once to
$\delta h_{hf}$ as ``a typical hyperfine energy'',
but make no use whatsoever of this non-definition.

We can similarly see no basis for the statement towards the end of the
Comment by ST \cite{ST}: ``... but here using the kinetic equation of 
Ref. [2], rather than
just a phenomenological noise.'', since the Monte Carlo simulations 
we perform in Ref.
\cite{ourPRL} happen to be \cite{exp} a stochastic implementation of 
``the kinetic
equation'', i.e., Eq. (4) of Ref.\cite{PSPRL}.

Finally, at the risk of being repetitious, we cannot understand the 
closing remark
in the preceeding Comment, ``...so we predict a steep fall of 
$\xi_0$, which would be very
hard to understand in the approach of ref. [4]''. Indeed, there is 
nothing to understand
about this in the approach of Ref. \cite{ourPRL}, because the physics 
of nuclear
baths which is behind TEW's is not at all within the scope of Ref. 
\cite{ourPRL}.

What we do show in Ref.\cite{ourPRL} is that a hole of half-width $W$
develops in the dipole field distribution of an Ising model of spins 
interacting through
dipolar fields with the flipping rule just mentioned, and that 
$W\simeq 0.75\delta
h_{hf}$, and we know of no previously published work establishing a 
relation between these two
quantities.
\rule{3in}{0.01in}

\noindent{Juan J. Alonso$^{1}$ and Julio F. Fern\'andez$^{2}$}\\
$^1${F\'{\i}sica Aplicada I, Universidad de M\'alaga,
29071-M\'alaga, Spain.}
$^2${ICMA, CSIC and Universidad de Zaragoza, 50009-Zaragoza, Spain}\\


\begin{thebibliography}{99}
\bibitem{ST}P.C.E. Stamp and I. S. Tupitsyn, arXiv:cond-mat/0211413.
\bibitem{ourPRL} J.J. Alonso and J.F. Fern\'andez, Phys. Rev. Lett. 
{\bf 87}, 097205 (2001).
\bibitem{PSPRL} N.V. Prokof'ev and P.C.E. Stamp, Phys. Rev. Lett., 
{\bf 80}, 5794 (1998).
\bibitem{PS1} N.V. Prokof'ev and P.C.E. Stamp, J. Low Temp. Phys. 
{\bf 104}, 143 (1996).
\bibitem{exp}Except that we use a square hat like $\tau^{-1}(\xi)$ 
instead of an exponential
function, and that Eq. (4) of Ref. \cite{PSPRL} only holds for in the 
$\delta h_{hf}/k_BT\rightarrow
0$ limit, in the notation of Ref. \cite{ourPRL}.

\end{thebibliography}
\end{document}